\documentclass[10pt,conference]{IEEEtran}
\IEEEoverridecommandlockouts
\newcommand{\dealabs}{\textbf{DEAL}}
\newcommand{\deal}{\textbf{\texttt{DEAL}}}
\newcommand{\qaoa}{\textbf{\texttt{QAOA}}}

\newcommand{\qpu}{\textbf{\texttt{QPU}}}

\usepackage{array} 
\usepackage{graphicx}  % For including figures
\usepackage[symbol]{footmisc}
\usepackage{svg}
\usepackage{adjustbox}
\usepackage{multirow}
\usepackage{subcaption}
\usepackage{tikz}
\usepackage{quantikz}      % quantum circuits
\usetikzlibrary{positioning,calc,shapes.multipart,arrows.meta,fit, quantikz2}

\tikzset{
operator/.append style={fill=red!20}, % Make gate background red
}

\usepackage{algorithm}
\usepackage{algorithmic}
\usepackage{physics}

\usepackage[T1]{fontenc}    % use 8-bit T1 fonts
\usepackage[utf8]{inputenc} % allow utf-8 input
\usepackage{hyperref}       % hyperlinks
\usepackage{url}            % simple URL typesetting
\usepackage{booktabs}       % professional-quality 
\usepackage{amsfonts}       % blackboard math symbols
\usepackage{nicefrac}       % compact symbols for 1/2, etc.
\usepackage{microtype}      % microtypography
\usepackage{xcolor}         % colors

\usepackage{stfloats} % Required for strip environment

\usepackage{amsmath,amsfonts,amssymb,amsthm,mathtools,nicefrac,mymath}

\usepackage{listings} % For code formatting
\usepackage{xcolor}   % Color customization

\usepackage[margin=1in]{geometry}
\usepackage{environ}     % unconnected control

\usepackage{booktabs}

\usepackage{cite}               % order multiple entries in \cite{...}
\usepackage{url}                % allow \url in bibtex for clickable links
\usepackage{xcolor}             % color definitions, to be use for...
\usepackage[]{hyperref}         % ...clickable refs within pdf...
\hypersetup{                    % ...like so
  colorlinks,
  linkcolor={green!80!black},
  citecolor={red!70!black},
  urlcolor={blue!70!black}
}
\usepackage[capitalize,nameinlink]{cleveref}

\usepackage{comment}

\usepackage{makecell}

\def\BibTeX{{\rm B\kern-.05em{\sc i\kern-.025em b}\kern-.08em
    T\kern-.1667em\lower.7ex\hbox{E}\kern-.125emX}}
\begin{document}

\title{Direct entanglement ansatz learning (DEAL) with ZNE on error-prone superconducting qubits\\
% \thanks{\url{https://github.com/gzquse/QUBO}}
}

\author{
\IEEEauthorblockN{Ziqing Guo\IEEEauthorrefmark{1}}
\IEEEauthorblockA{
\textit{Texas Tech University} \\
Lubbock, USA \\
ziqguo@ttu.edu}
\and
\IEEEauthorblockN{Steven Rayan}
\IEEEauthorblockA{\textit{University of Saskatchewan} \\
Saskatoon, Canada \\
rayan@math.usask.ca}
\and
\IEEEauthorblockN{Wenshuo Hu}
\IEEEauthorblockA{\textit{Texas Tech University} \\
Lubbock, USA \\
wenshuo.hu@ttu.edu}
\and
\IEEEauthorblockN{Ziwen Pan}
\IEEEauthorblockA{\textit{Texas Tech University} \\
Lubbock, USA \\
ziwen.pan@ttu.edu}
\thanks{\IEEEauthorrefmark{1}Also affiliated with NERSC, Lawrence Berkeley National Laboratory, Berkeley, USA.}
}

\maketitle

\begin{abstract}
    Quantum combinatorial optimization algorithms typically face challenges owing to complex optimization landscapes featuring numerous local minima, exponentially scaling latent spaces, and susceptibility to quantum hardware noise. In this study, we introduce Direct Entanglement Ansatz Learning (DEAL), wherein we employ a direct mapping from quadratic unconstrained binary problem parameters to quantum ansatz for cost and mixer Hamiltonians, which improves the convergence rate towards the optimal solution. Our approach exploits a quantum entanglement-based ansatz to effectively explore intricate latent spaces and zero-noise extrapolation (ZNE) to mitigate the randomness caused by crosstalk and coherence errors. Our experimental evaluation demonstrates that \deal\ increases the success rate compared to the classic quantum approximation optimization algorithm while controlling the error variance. In addition, we demonstrate the capability of DEAL to provide near-optimum ground energy solutions for traveling salesman, knapsack, and maxcut problems, which facilitates novel paradigms for solving relevant NP-hard problems and extends the self-contained practical applicability of quantum optimization using noisy quantum hardware. 
\end{abstract}

\section{Introduction}
% qubo 
Quantum approximate optimization algorithms (QAOA) and their derivatives constitute a class of methodologies aimed at addressing combinatorial optimization (CO) problems, which involve identifying optimal configurations within a discrete yet extensive search space. Consequently, these techniques have been applied across various domains, including logistics \cite{dalal2024digitized}, global carbon emission management \cite{climate}, quantum cryptography \cite{khurana2022applications}, and atom-level analysis \cite{wang2024atomique}. Furthermore, significant advancements have facilitated the development of innovative approximation strategies that utilize superconducting quantum processors, thereby establishing them as pivotal mechanisms for quantum-enhanced computational paradigms \cite{sachdeva2024quantum}.

Classical methodologies, including grid search \cite{arora2003approximation}, tabu search \cite{glover1998adaptive}, and Markov Chain Monte Carlo \cite{hochba1997approximation}, have shown potential in approximating NP-hard problems, such as the knapsack problem (KP) \cite{martello1987algorithms}, MaxCut \cite{ben2014maximum}, and the traveling salesman problem (TSP) \cite{hoffman2013traveling}. However, their efficacy is inherently limited by the binary nature of classical computing methods. This dependence on definitive 0s and 1s presents substantial challenges in solving combinatorial optimization (CO) problems, as it necessitates the conventional quadratic unconstrained binary optimization (QUBO) formulation, which inherently results in data structures that contribute to an exponential increase in computational complexity \cite{kannan2005approximation}. In this context, the quantum formulation of the Ising model~\cite{cipra1987introduction} maps the system to a high-dimensional Hilbert space, which can be efficiently explored using variational and gate-based quantum circuits~\cite{biamonte2021universal, cerezo2021variational}. These circuits incorporate unitary operations that evolve over time~\cite{du2022quantum, schuld2020circuit}.

% details of quantum problem 
Specifically, the energy landscape of the Quantum Approximate Optimization Algorithm (QAOA) frequently displays pervasive local minima owing to the limited search capacity inherent in quantum circuits, which is contingent upon the construction of the ansatz~\cite{mcclean2018barren}. These limitations arise from constraints associated with the initialization of parameterized quantum gates, restricted connectivity within quantum processing units (QPUs), and the number of feasible entanglement and unitary gates, also known as circuit depth. Consequently, the QAOA is highly sensitive to the quantum circuit ansatz, leading to more challenging optimization processes and a limited practical advantage over classical methods of quantum computing. For a comprehensive understanding, we refer to the work of \cite{sack2021quantum} regarding annealing-type QPU ansatz construction; however, in this study, we focus on quantum circuit construction utilizing a superconducting-based QPU.

% contribution
This study introduces the Direct Entanglement Ansatz Learning (DEAL), a method designed to enhance QAOA learning on IBM's noisy superconducting hardware. The \deal\ framework extends the classical QUBO formulation by mapping its variables onto a cost Hamiltonian derived from the objective function, which is optimized by qubit connectivity. This approach encodes the problem constraints and defines expected outcomes. Furthermore, the method incorporates zero-noise extrapolation (ZNE)~\cite{zne} to mitigate the crosstalk noise inherent in superconducting quantum systems~\cite{arute2019quantum}. The structure of this paper is as follows: \cref{sec: background} provides a brief overview of QAOA and its derivatives. \cref{sec:med} introduces the enhanced parameterization learning, noise reduction strategies for the scheme, and dynamic metrics. \cref{sec:res} presents the empirical results of circuit expressivity evaluation concerning our model, the interpretability analysis of realistic superconducting qubits transpilation in the context of the current noise intermediate-scale quantum (NISQ) era, and numerical shots-based quantum circuit analysis. The notations used in this paper are summarized in \cref{tab:notation}.

% By doing so, it effectively expands the search space capacity of parameterized quantum circuits (PQCs), leading to improved simulation stability, higher success rates, and faster convergence rates.

% remove verbose details:
% utilizing bayesian optimization with adaptive polynomial regression \cite{polynomial} and a Gaussian process \cite{gaussian}. 

\begin{comment}

We propose the  that leverages global optimization for the cost Hamiltonian (that is, Bayesian optimization with adaptive polynomial regression \cite{polynomial} and a Gaussian process \cite{gaussian}), direct parameters mapped non-local crosstalk parameterized quantum circuit anstaz (PQC), and zero noise extrapolation (ZNE) \cite{zne}, which exploit the search space capacity of PQC for QAOA-based framework in regards of simulation stability, success rate, and convergence time (see details in \cref{sec:res} and \cref{med}). We demonstrate the generality of QCOA by processing the QUBO problem and improving the performance in TSP, KP, and Maxcut problems. Furthermore, we mark the threshold of current \qpu limitation for QCOA concerning IBM fast-evolving backends (e.g. 133 qubit Torino and Marakash)~\footnote{https://quantum.ibm.com/services/resources} after implementing the mitigation techniques. 

\end{comment}
% the sequence might need to be adjust 
\section{Background}
\label{sec: background}
We refer the QAOA quantum learning paradigm in the original work \cite{farhi2014quantum}, improved by dynamic QAOA models \cite{zhu2022adaptive, cheng2024quantum} and QAOA-GPT \cite{tyagin2025qaoa
}. 
Given a QUBO instance  
\begin{equation}
    \max_{x\in\{-1,1\}^{n}}\;x^{\mathsf T}Wx=\sum_{i<j}w_{ij}x_i x_j , \quad
 x_i\in\{-1,1\},
\label{eq:qubo}
\end{equation}
whose special cases include MaxCut, KP and TSP, one encodes the objective into an Ising-type cost Hamiltonian on $n$ qubits,
\begin{equation}
   H_C \;=\; -\tfrac12 \sum_{i<j} w_{ij}\bigl(I-Z_i Z_j\bigr), 
   \label{eq:qubocost}
\end{equation}
where $Z_i$ is the Pauli–$Z$ operator acting on qubit $i$.  The QAOA prepares, for a chosen depth $p$, the variational state  
\begin{align}
|\psi(\boldsymbol{\gamma},\boldsymbol{\beta})\rangle
      &=\Biggl[\prod_{\ell=1}^{p} 
               e^{-i\beta_\ell H_B}\,
               e^{-i\gamma_\ell H_C}\Biggr]
        |+\rangle^{\otimes n},
        \label{eq:qaoa_state}\\[2pt]
H_B   &=\sum_{i=1}^{n} X_i.
        \label{eq:qaoa_mixer}
\end{align}
The classical optimizer adjusts the angles within $\{\gamma_\ell,\beta_\ell\}$ to minimize $\langle\psi|H_C|\psi\rangle$. Adaptive variants, such as ADAPT-QAOA \cite{zhu2022adaptive} iteratively grow the operator pool to suppress barren plateau effects, while QAOA-GPT \cite{tyagin2025qaoa} treats the ordered gate and angle list as a token sequence and utilizes a pre-trained transformer to predict good updates. Both strategies alleviate the problem but still have acute classical search overhead because parameter tuning is merely shifted to gradient estimation (ADAPT) or to a language model that still needs fine-tuning.
Following the spirit of few-shot QAOA \cite{hao2024end} and its prior work \cite{sureshbabu2024parameter}, we present the \deal\ scheme to further improve the QAOA-like computation paradigm. 

The implementation of the \deal\ model, built upon openQAOA and qiskit, is publicly available at \url{https://github.com/gzquse/DEAL\_QUBO}. Using state-of-the-art on-demand superconducting QPUs and high-performance statevector simulators, we assume throughout this paper that the default 1,024 shots per quantum circuit yield plausible results in the NISQ era.

\begin{table}[t]
\centering
\caption{Notation}
\begin{tabular}{cl}
\toprule
\textbf{Symbol} & \textbf{Description} \\
\midrule
$n$ & Number of binary variables (qubits) \\
$p$ & QAOA depth (layer index $k = 1,\ldots,p$) \\
$Q \in \mathbb{R}^{n \times n}$ & Upper-triangular QUBO matrix, \\
& off-diagonal elements $Q_{ij}$ ($i<j$) \\
$\theta$ & QAOA parameters $\theta \coloneqq (\gamma_1,\ldots,\gamma_p, \beta_1,\ldots,\beta_p)$ \\
\bottomrule
\end{tabular}
\label{tab:notation}
\end{table}
\begin{figure*}
    \centering
    \includegraphics[width=1.0\linewidth]{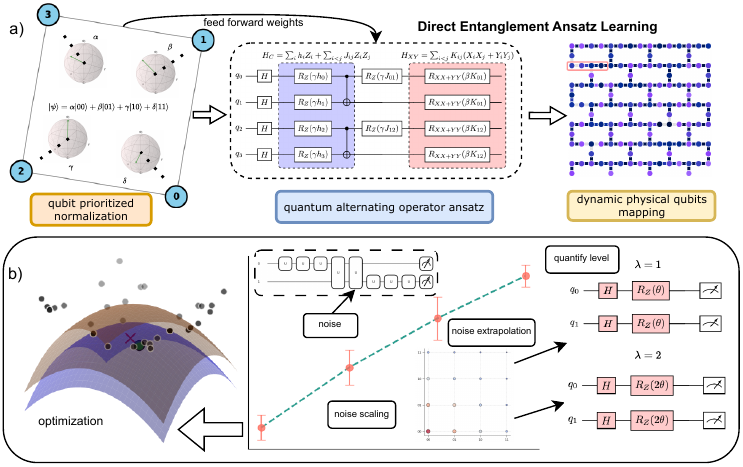}
    \caption{The proof-of-concept workflow for \deal.\ involves the QPN utilizing objective weights from the QUBO to the XY mixer Hamiltonian, followed by mapping to the real QPU topology. The variable \(\gamma\) represents the single rotation angles, whereas \(\beta\) denotes the controlled rotation angles, which range from 0 to \(\pi\). Additionally, the adaptive ZNE with Pauli gate insertion between the quantum ansatz and the rotation is regulated by the noise level \(\lambda\), which is adjusted by the classical optimizer.}
    \label{fig:med}
\end{figure*}

\section{Model}
\label{sec:med}
\subsection*{Direct entanglement ansatz learning (DEAL)}
\label{med:deal}
We demonstrate our model in \cref{fig:med}, with the top panel showing the \deal\ pattern followed by ZNE for noisy QPU optimization, as detailed in \cref{alg:deal}. To improve the convergence rate performance, we propose qubit-prioritized normalization (QPN), indicated by the first component in a). The quantum gate rotation angles depend on the number of cost Hamiltonian layers specified in the quantum alternating operator ansatz \cite{hadfield2019quantum} encoder.
\begin{algorithm}[h]
\caption{DEAL with Dynamic Physical Mapping}
\begin{algorithmic}[1]
\REQUIRE QUBO matrix $Q$, depth $p$, device topology $G_{device}$, error rates $E_{ij}$
\ENSURE optimized $\theta^*$, physical mapping $\pi$

\STATE \textbf{QPN:} Compute $s_i = \sum_j |Q_{ij}|$, normalize $w_i = s_i/\sum_\ell s_\ell$
\STATE Initialize $\gamma_k^0 = \frac{\lambda_\gamma k}{np} \sum_i \arccos(1-2w_i)$, $\beta_k^0 = \frac{\lambda_\beta(p-k)}{np} \sum_i \arcsin(\sqrt{w_i})$

\STATE \textbf{Mapping:} Sort qubits by $w_i$, minimizing $\sum_{ij} w_{ij} \cdot d(\pi(i),\pi(j)) \cdot E_{\pi(i)\pi(j)}$

\STATE \textbf{Ansatz:} Set $\theta^0 = (\gamma_1^0,\ldots,\gamma_p^0, \beta_1^0,\ldots,\beta_p^0)$
\FOR{$t = 1$ to optimizer budget}
    \STATE Evaluate $f(\theta_t) = \langle \psi(\theta_t) | H_C | \psi(\theta_t) \rangle$ with $S$ shots
    \STATE Update $\theta_{t+1} \leftarrow \text{OptimizerStep}(\theta_t, f(\theta_t))$
\ENDFOR

\RETURN $\theta^*$, $\pi$
\end{algorithmic}
\label{alg:deal}
\end{algorithm}

\begin{figure}
    \centering
    \includegraphics[width=1.0\linewidth]{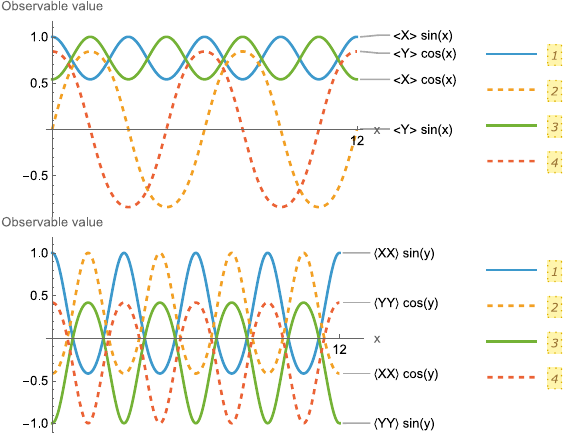}
    \caption{\textbf{Top:} Solid lines represent observable values for Rz rotation cost Hamiltonian with X and Y bases. \textbf{Bottom:} XY mixer Hamiltonian observables are encoded using sin function (solid lines). Dashed lines indicate encoding methods not employed in our experiment.}
    \label{fig:obs}
\end{figure}

The QPN procedure consists of three sequential steps. First, we compute the raw importance score per qubit for the QUBO problems as follows:
\begin{equation}
s_i = \sum_{j=1}{n} |Q_{ij}|.
\end{equation}
Dynamic physical qubit mapping enables the pre-selection of optimal superconducting qubits with minimal error rates, which has been proven to be state-of-the-art through the SABRE layout \cite{li2019tackling}. Second, we normalize these scores to obtain a probability distribution
\begin{equation}
w_i = \frac{s_i}{\sum_{\ell=1}{n} s_\ell},
\label{eq:dis}
\end{equation}
where $0 \leq w_i \leq 1$ and $\sum_i w_i = 1$. Third, our mixer Hamiltonian encoding method employs angle encoding, in which each observable state represents the importance of the problem variables. For $p$ layers, \deal\ parameterizes variables into each unitary operation. For every layer $k = 1,\ldots,p$, we define two angle tensors as follows:
\begin{align}
    \Phi^{\gamma}_{k,i} &= \lambda_\gamma \cdot \frac{k}{p} \cdot \arccos(1-2w_i), 
    \label{eq:cost} \\
    \Phi^{\beta}_{k,i} &= \lambda_\beta \cdot \left(1-\frac{k}{p}\right) \cdot \arcsin(\sqrt{w_i}),
    \label{eq:mixer}
\end{align}
where $\lambda_\gamma, \lambda_\beta \in (0,\pi]$ are global scaling factors chosen per problem class ($\lambda_\gamma=\pi$, $\lambda_\beta=\nicefrac{\pi}{2}$ by default).
\begin{equation}
    \centering
    \begin{quantikz}
    \lstick{$q_x$} & \qw & \gate{H} & \gate{R_Z(\theta)}  & \meter{} \\
    \lstick{$q_y$} & \gate{S^\dag} & \gate{H} & \gate{R_Z(\theta)} & \meter{}
    \end{quantikz}
    \label{circ}
\end{equation}
 Although $q_y$ in circuit \eqref{circ} provides <Y> measurement basis for the phase representation, the QPN employs the $q_x$ pattern for the cost Hamiltonian to ensure ZNE compatibility through Pauli gate insertion. The <X> basis in $q_x$ (X=HZH \cite{mitarai2021constructing}, Z:=Rz($\pi$)) demonstrates a smaller amplitude than the <Y> basis shown in \cref{fig:obs}, resulting in enhanced noise resilience. We defer the analysis of the QPN in \cref{app: qPN}.
Large QUBO matrix values $|Q_{ij}|$ or $|Q_{ii}|$ indicate qubit $i$ importance, affecting $s_i$ and $w_i$. The layer index $k$ ensures that the early layers emphasize cost unitaries (larger $\Phi_\gamma$), whereas the later layers prioritize mixing (larger $\Phi_\beta$). Nonlinear trigonometric encodings prevent saturation near $0$ or $\pi$. \deal\ transforms matrices from \cref{eq:qaoa_mixer} into scalar layer parameters by averaging over qubits, yielding initial parameters shown in \cref{eq:avg_init}
\begin{equation}
    \gamma_k^0 = \frac{1}{n} \sum_{i} \Phi^{\gamma}_{k,i}, \quad \beta_k^0 = \frac{1}{n} \sum_{i} \Phi^{\beta}_{k,i}.
    \label{eq:avg_init}
\end{equation}
Fixed initialization provides zero variance but ignores the problem structure, while random initialization exhibits high variance for uniform distribution \cite{sureshbabu2024parameter}
\begin{equation}
   Var[\theta^0] = \frac{(2\pi)^2}{12}.
\label{eq:Var}
\end{equation}
Our initialization achieves an optimal bias-variance trade-off by maintaining a low variance through deterministic computation while incorporating problem-specific bias from \cref{eq:avg_init}. We also refer to three parameter initialization cases in \cref{app:cases}. The early layers emphasize exploration (larger $\beta_k^0$ from $(1-k/p)$ weighting), whereas the later layers emphasize exploitation (larger $\gamma_k^0$ from $k/p$ weighting). We derive the expected distance from \cref{eq:avg_init} and \cref{eq:Var}
\begin{equation}
\mathbb{E}[\|\theta^0_{\text{QPN}} - \theta^*\|] \leq \mathbb{E}[\|\theta^0_{\text{random}} - \theta^*\|].
\label{eq:distance}
\end{equation}
Nonlinear mappings $\arccos(1-2w_i)$ and $\arcsin(\sqrt{w_i})$ prevent parameter saturation at boundaries $\{0, \pi\}$ where gradients vanish, maintaining non-zero gradient flow throughout the optimization and avoiding gradient vanishing problems in poorly initialized quantum circuits \cite{mcclean2018barren}.

\subsection*{Adaptive ZNE for \dealabs}
\label{med:zne}
Increasing layer count exacerbates the over-rotation and crosstalk issues, resulting in poor initial guesses and slow convergence rates. Because quantum noise can be numerically estimated and measured at the gate level, ZNE provides noise-free observable expectation values by analyzing expectation values across varying noise intensities \cite{kandala2019error, PhysRevLett.119.180509}. Although surface codes enable physical noise stability \cite{preskill2025beyond}, we implemented ZNE for IBM QPUs to enable scalable operations with minimal overhead and no ancilla qubit requirements.

As specified in the dashed box of (b) in \cref{fig:med}, noise cancellation employs Pauli twirling and gate-flip correction. \deal\ iteratively formulates and refines quantum circuit outputs through Bayesian posterior updates of noise scaling parameters, where each factor is defined as $\lambda_{ij} = w_i \cdot w_j \cdot d_{ij}$ with $w_i, w_j$ representing QPN importance weights from normalized QUBO row sums and $d_{ij}$ denoting qubit distance in the coupling map. Note that $\lambda$ is rounded to integer values (proof deferred to \cref{app: baye}). Rather than classical polynomial fitting for noise extrapolation (unitary folding and parameterized noise scaling \cite{giurgica2020digital}), our approach leverages QPN-weighted distance metrics to optimize qubit connectivity by prioritizing high-importance qubits in the controlled gate allocation and circuit routing. Iterative Bayesian refinement is defined as
\begin{equation}
   \boldsymbol{\lambda}^{(t+1)} = \boldsymbol{\lambda}^{(t)} + \boldsymbol{\Sigma}_{\text{prior}} \mathbf{J}^T (\boldsymbol{\Sigma}_{\text{noise}})^{-1} \boldsymbol{r}. 
\label{eq:lambda}
\end{equation}
This minimizes the connectivity cost $C_{\text{conn}} = \sum_{i<j} w_i w_j d_{ij} n_{ij}$, ensuring that critical qubits identified by the QPN analysis receive preferential treatment in error correction and circuit compilation, yielding noise-mitigated expectation values weighted by the QUBO problem structure. Since coupling maps evolve due to real-world noise, we select consecutive qubits as the minimum error-rate coupling map\footnote{More information in qiskit BackendV2 coupling map: \url{https://docs.quantum.ibm.com/api/qiskit/qiskit.providers.BackendV2}}.
% motivation and detials

\subsection*{Dynamic metrics}
\label{med: metrics}
We define the quantum noise-limited relative error (QNRE) metric to characterize the final optimization results by problem type. QNRE quantifies the extent to which quantum ansatz approximates near-optimal solutions within noise constraints
\begin{equation}
    \text{QNRE} = \frac{E_{\text{observed}} - E_{\text{optimal}}}{E_{\text{optimal}}}.
    \label{eq: qnre}
\end{equation}
QNRE is intrinsically correlated with the global energy landscape. Given the observed energy \(E_{\text{observed}}\) and optimal energy \(E_{\text{optimal}}\) (ground state energy), the metric determines whether quantum noise obstructs optimal solution attainment.
When the deviation satisfies \( |E_{\text{observed}} - E_{\text{optimal}}| \leq E_{\text{noise}} \), quantum noise significantly overshadows the optimal energy. We provide the general function
\begin{equation}
    \text{QNRE}_{j}^{(i)} = 
    \frac{\max\left(\left|E_{\text{observed},j}^{(i)} - E_{\text{optimal}}^{(i)}\right| - E_{\text{noise}}^{(i)},\, 0\right)}{E_{\text{optimal}}^{(i)}} ,
    \label{eq: qnre_noise}
\end{equation}
where \(i\) represents the problem index, and \(j\) denotes the eigenvalue.

\begin{figure}[ht]
    \centering
    \includegraphics[width=1.0\linewidth]{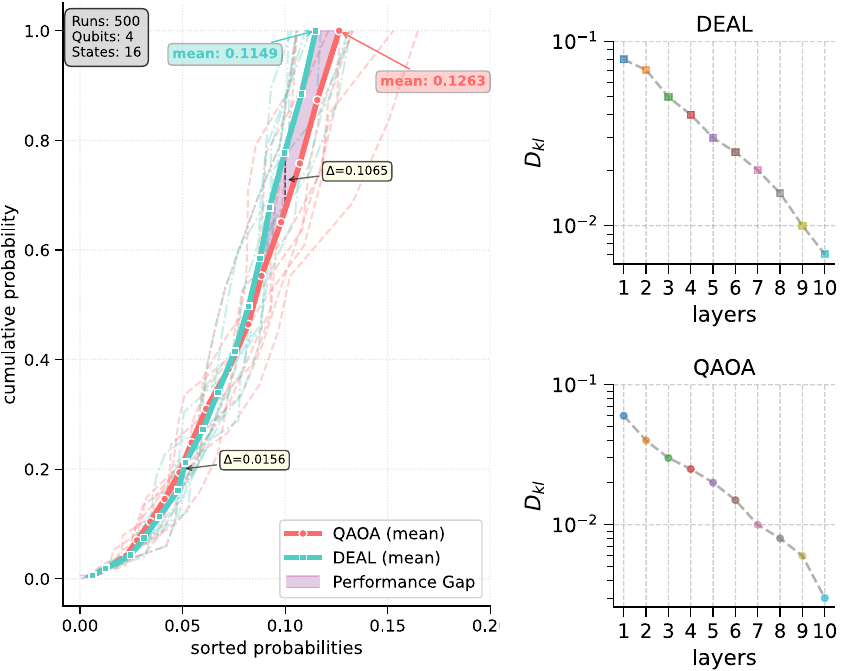}
    \caption{\textbf{(Left)} cumulative distribution functions (CDFs) of measurement outcome probabilities for \deal\ and QAOA algorithms across 500 independent runs on a 4-qubit complete graph (graph refer in \cref{fig:med}). Denote that the individual runs averaged over 50 batches are shown as semi-transparent dashed lines, with thick lines representing the mean performance across all runs. \textbf{(Right)} Kullback-Leibler (KL) divergence between algorithm output distributions and uniform Haar-random distributions as a function of circuit depth (1-10 layers). We refer the benchmark details in \cref{app:benchmark}.} 
    \label{fig:res1}
\end{figure}

\section{Results}
\label{sec:res}
% problems
This section outlines the principal experiments, covering the circuit search capabilities, quantum computer transpilation interpretability, and NP-hard problem analysis.

\subsection*{Circuit capacity examination}
\label{res: cap}
In our study, we first evaluated the computational capability of the \deal\ framework using Erdos--Renyi graph (ERG) with MaxCut task \cite{erdds1959random}, where each edge is created with a 20\% probability; we note that such settings allow control graph density and connectivity, ensuring that the resulting non-fully connected graphs introduce a more complex search space for the quantum ansatz.
\cref{fig:res1} illustrates that the framework provides a more evenly distributed probability concentration, as indicated by the shaded area.
Furthermore, the generality of our approach becomes clearer with an increasing number of layers in the framework, indicating that the vanilla QAOA is more likely to suffer from the barren plateau issue but with more expressivity power. 
This is nontrivial because QPN prioritizes the qubit coupling map based on the problem graph layout as formulated in \cref{eq:distance}, and also uses ZNE postprocessing to numerically mitigate the randomize expressivity of the latent space.

\begin{figure}
    \centering
    \includegraphics[width=0.62\linewidth]{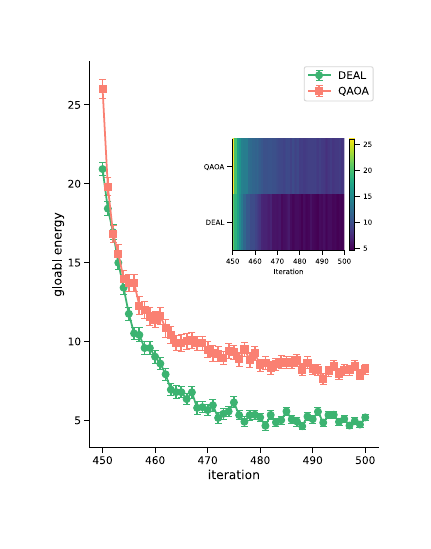}
    \caption{The figure shows the detail of convergence rate across the last 50 runs between \deal\ and QAOA with both error bars are decreasing (note that we use single layer to demonstrate the low parameters setting performance). We denote the inset that more darker color symbolizes better optimization with lower cost.}
    \label{fig:converge}
\end{figure}

The benefits become evident when evaluating lower-depth circuits, as shown in \cref{fig:converge}. The pre-parameterized cost Hamiltonian derives from \cref{eq:cost}
\begin{equation}
\begin{aligned}
    H_C &= \sum_{i} \left( \frac{Q_{ii}}{2} + \sum_{j \neq i} \frac{Q_{ij}}{4} \right) 
    - \sum_{i} \frac{Q_{ii}}{2} Z_i \\
    &\quad - \sum_{i<j} \frac{Q_{ij}}{4} (Z_i + Z_j) 
    + \sum_{i<j} \frac{Q_{ij}}{4} Z_i Z_j,
\end{aligned}
\label{eq: Hc}
\end{equation}
which re-expresses the problem coefficients \(Q_{ij}\) as Pauli-Z interactions. Treating graph connections as XY mixer Hamiltonian \cite{wang2020xy} qubit connectivity yields
\begin{equation}
    H_M = \sum_{(i,j) \in E(G(n, p))} \frac{1}{2} \left( X_i X_j + Y_i Y_j \right),
    \label{eq: Hm}
\end{equation}
where \( G(n, p) \) represents the ERG with \( n \) qubits and edge probability \( p \). \cref{eq: Hc} and \cref{eq: Hm} stabilize quantum state distributions before applying learning Hamiltonian and enhance resource efficiency by eliminating redundant layers. Reducing the circuit depth extends the coherent evolution time in noisy environments, effectively mitigating decoherence effects.

\begin{figure}
    \centering
    \includegraphics[width=0.9\linewidth]{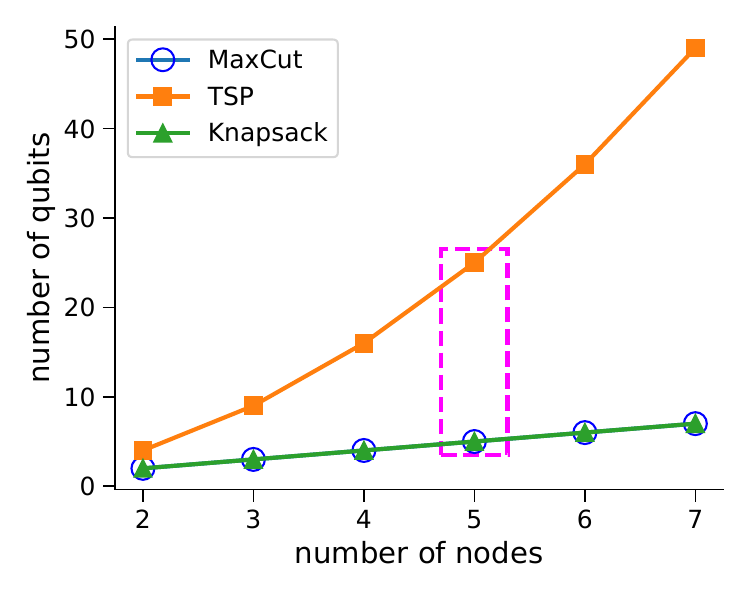}
    \caption{The number of qubits requirement as the increasing with the problem size shown respectively by maxcut, travel salesman, knapsack problems. The rectangle mark represent the typical five nodes problem experimented for hardware and numerical analyses.}
    \label{fig: probs}
\end{figure}
\subsection*{Physical hardware evaluation}
\label{res:trans}
% reason
\begin{figure}
    \centering
    \includegraphics[width=0.75\linewidth]{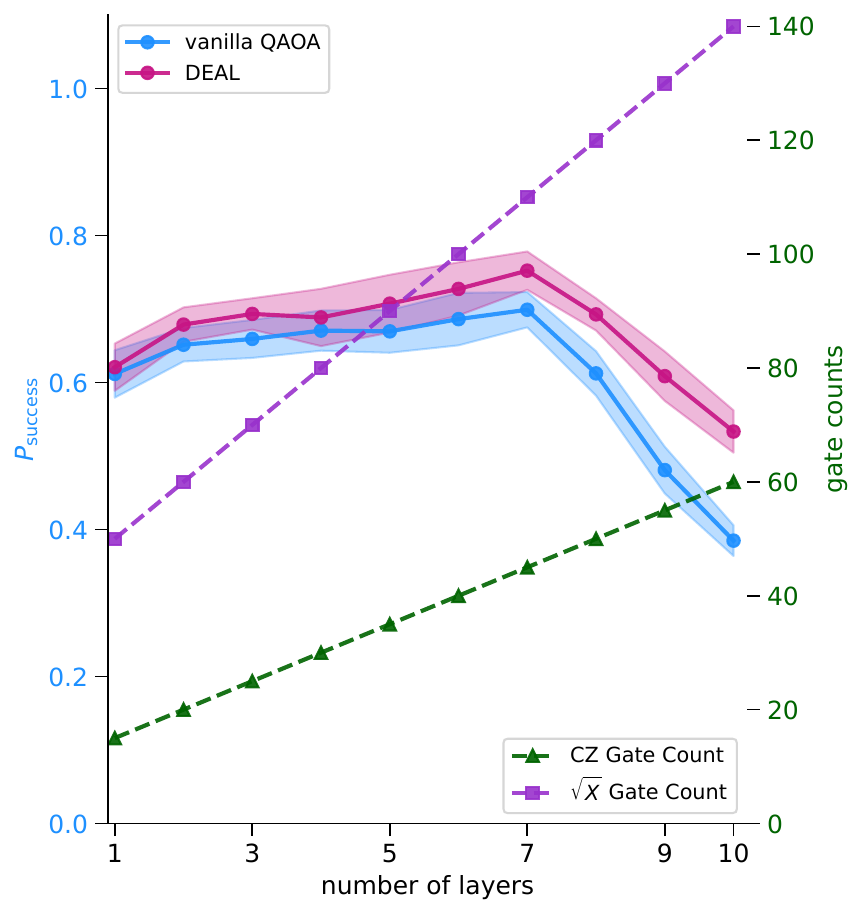}
    \includegraphics[width=0.75\linewidth]{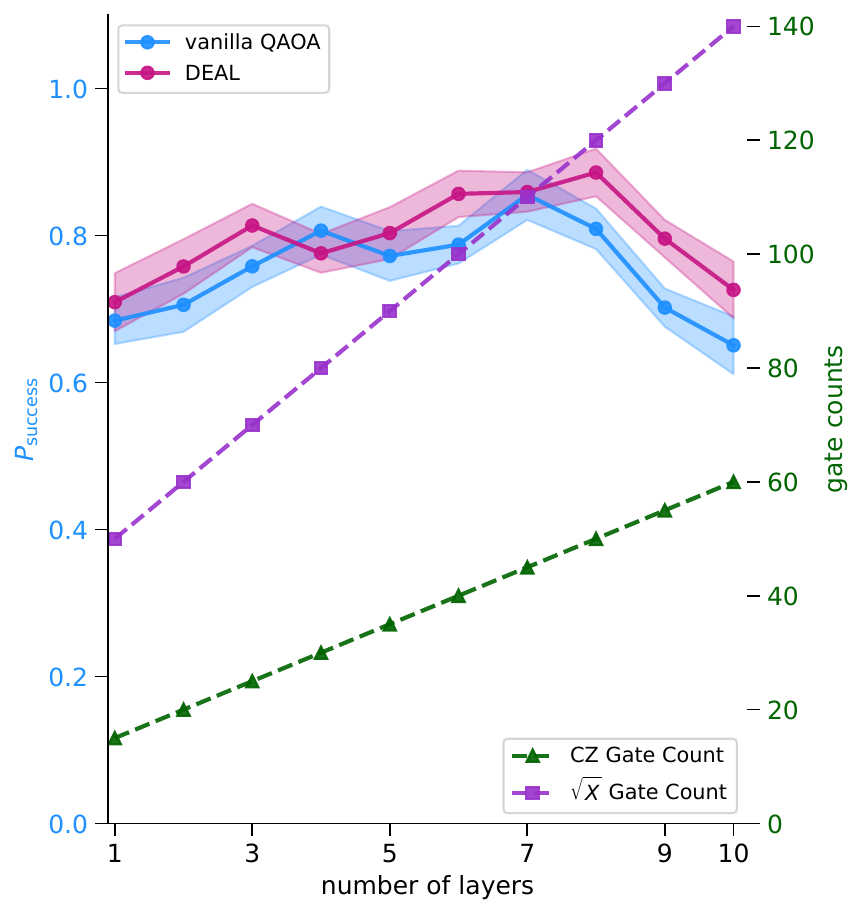}
    \caption{Performance comparison of vanilla \qaoa\ and \deal\ algorithms executed on IBM 133-qubit Torino \textbf{(above)} and 156-qubit Marrakesh \textbf{(below)}. The shaded area indicates the error rate.}
    \label{fig:res2}
\end{figure}

In a broader impact scenario, superconducting quantum computing has evolved rapidly \cite{hua2015fast, acharya2024quantum, sanders2025superconducting}. We note that the total number of CZ gates and square root X operations constrain the computational accuracy, as excessive entangling operations and frequent measurements degrade the coherence and introduce errors \cite{chen201864, chen2018classical}.

\begin{table}[ht]
\centering
\caption{Success rate differences for Torino and Marrakesh across 10 layers.}
\begin{tabular}{c c c}
\toprule
\multirow{2}{*}{$p$ (layer)} & \multicolumn{2}{c}{success rate difference (\%)} \\
\cmidrule(lr){2-3}
            & \textbf{Torino} & \textbf{Marrakesh} \\
\midrule
1  &  0.91 &  2.51 \\
2  &  2.74 &  5.22 \\
3  &  3.40 &  5.55 \\
4  &  1.79 &  -3.09 \\
5  &  3.77 &  3.15 \\
6  &  4.11 &  6.90 \\
7  &  5.32 &  0.39 \\
8  &  8.05 &  7.69 \\
9  & 12.80 &  \textbf{9.30} \\
10 & \textbf{14.81} &  7.40 \\
\bottomrule
\end{tabular}
\label{tab:success_difference}
\end{table}

Here, we select a typical five-vertex fully connected graph as qubits and nodes, indicated by the magenta rectangle in \cref{fig: probs}. \deal\ outperforms vanilla QAOA with over 14\% higher success rate in finding ground-state energy using Heron-type QPUs, as shown in \cref{fig:res2} and detailed in \cref{tab:success_difference}. Noise perturbations significantly impact performance beyond approximately 43 and 41 CZ gates in Torino and Marrakesh, respectively, resulting in 20\% measurement outcome randomness despite the ZNE mitigation techniques.

We note that the standard deviation gradually decreases as the quantum circuit simulation evolves, indicating improved optimization convergence before implementing the seven layers. However, superconducting qubit decoherence limitations, characterized by \(T_1\) and \(T_2\) relaxation times \cite{clarke2008superconducting, devoret2013superconducting}, and readout error rates \cite{aasen2024readout} constrain performance. Current state-of-the-art QPUs struggle with increasing circuit depths and non-local gate presence.
Specifically, in ZNE, delay gates \cite{delay} function as identity gates inserted within the compiled circuits. Increasing the quantum processing time enhances noise mitigation before \deal\ reaches the seven-layer bottleneck, where noise dominates. The Echoed Cross-Resonance (ECR) gate \cite{tripathi2019operation}, defined as \(\frac{1}{2} (IX - XY)\), provides phase mitigation in higher excited states while generating maximal entanglement, improving performance beyond the seven-layer depth. Our approach maintains the variance within reasonable ranges while enhancing the ground truth search capability. As shown in \cref{tab:success_difference}, the method provides more efficient dynamic ansatz encoding through the Torino QPU.

\begin{figure*}[ht]
    \centering
    \includegraphics[width=0.31\linewidth]{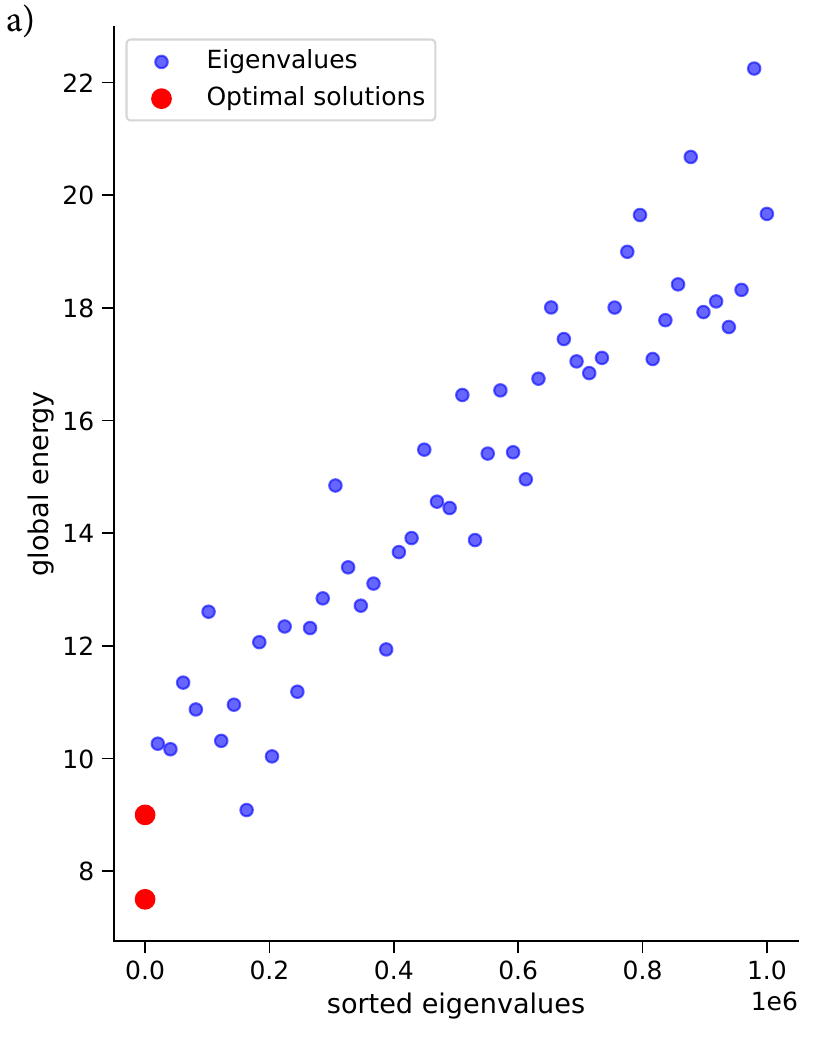}
    \includegraphics[width=0.31\linewidth]{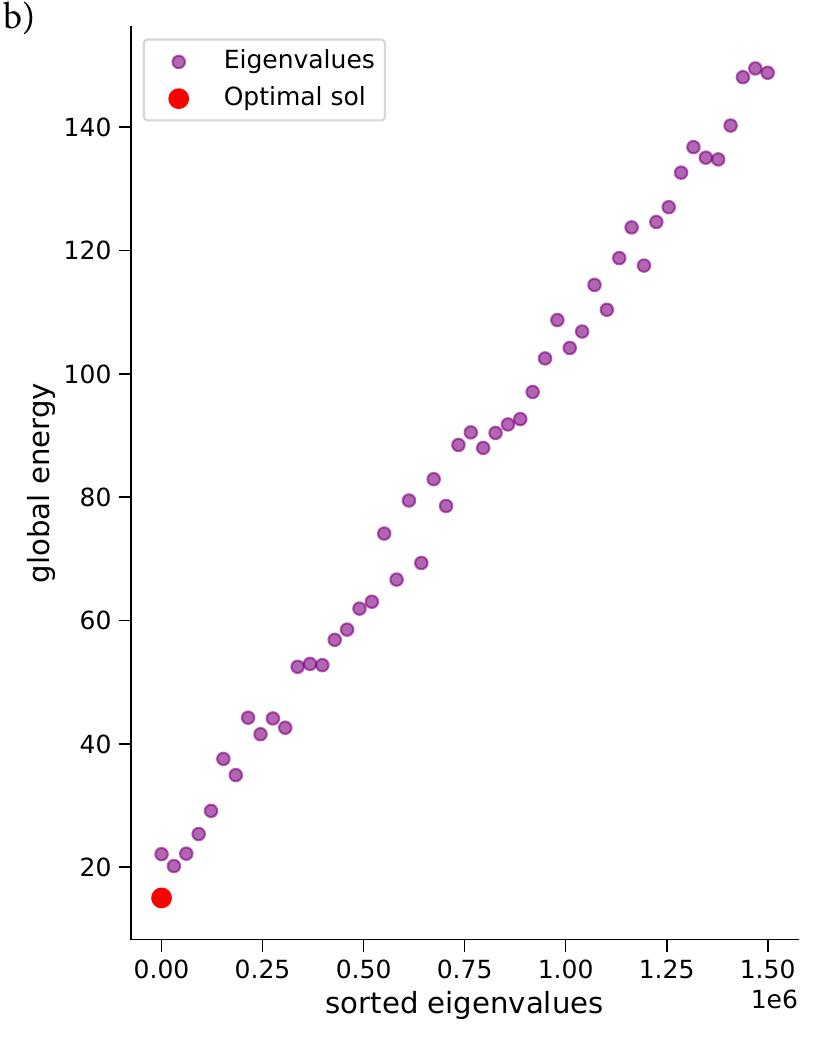}
    \includegraphics[width=0.31\linewidth]{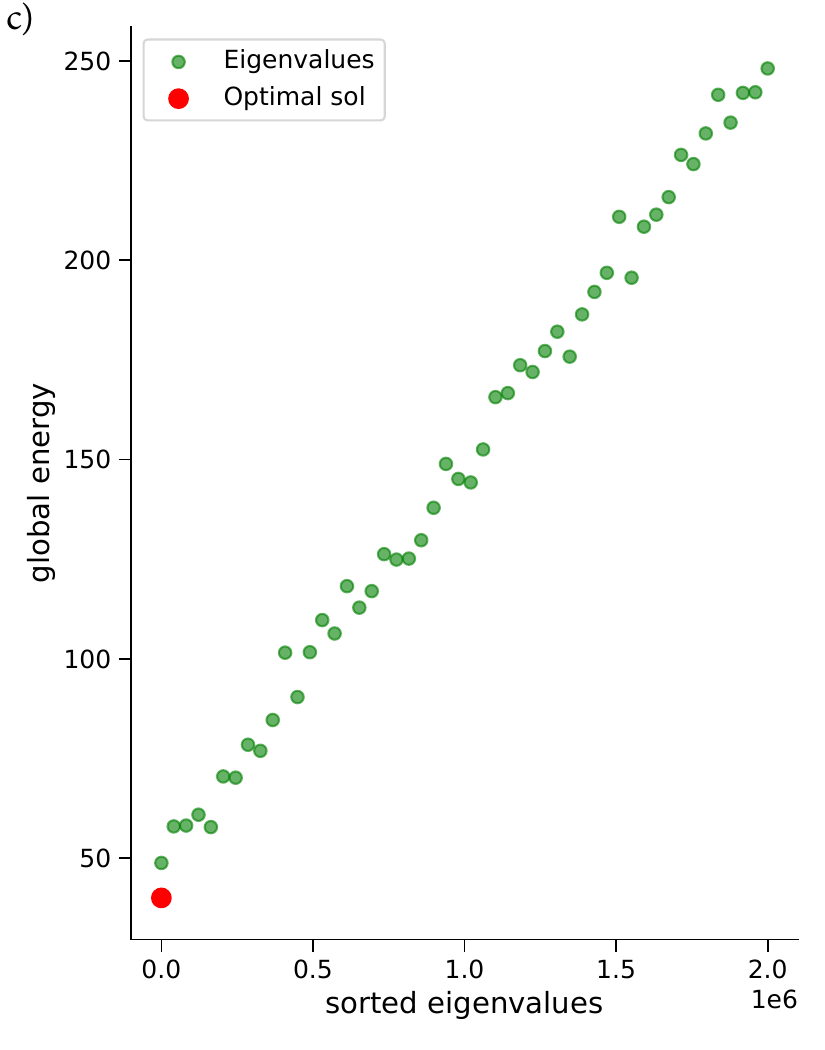}
    \caption{Numerical simulation conducted on ideal statevector simulator with default 1,024 shots utilizing \deal\ paradigm with 50 best selected eigenvalues, where the problems are a) TSP, b) MaxCut, c) KP and red dots mark as the optimum solutions with the lowest eigenvalue.}
    \vspace{-10pt}
    \label{fig:res3}
\end{figure*}

\subsection*{Problem numerical analysis}
\label{res:probs}
We selected three typical NP-hard problems to validate the versatility of the \ deal \ paradigm. In \cref{fig:res3}, our method successfully achieved a near-optimal ground-state energy across all three problems.
Smaller eigenvalues in TSP encoding result in reversed ground-state energy due to two-way graph utilization, reflecting multiple valid solutions in non-Eulerian graphs \cite{fleischner1990eulerian} after the final measurement. The eigenvalue units are defined as \(10^6\), where the QUBO matrix objective values sampled by the shot-based simulator contribute to the energy landscape eigenvalues.
We observe that our approach evades inefficient local minima by extracting and removing uncorrelated noise at different levels, as illustrated by the quasi-linear distributions in panels b and c of \cref{fig:res3}.

\begin{table}[htbp]
\centering
\caption{The QNRE metric for selected 50 eigenvalues.}
\begin{tabular}{ccc}
\toprule
 \textbf{Problem}&\textbf{Energy range} & \textbf{QNRE range (\%)} \\ 
 \midrule
 TSP & $7$--$22$ & $7.7\text{–}69$ \\[2pt]
 MaxCut & $50\text{–}250$ & $25\text{–}550$ \\[2pt]
 KP & $18\text{–}140$ & $20\text{–}833$ \\[2pt]
\bottomrule
\end{tabular}
\label{tab:qnre}
\end{table}
However, this is not the primary factor in our scenario, since MaxCut and KP require two-qubit entanglement, unlike TSP, which involves multi-qubit entanglement. The QNRE metric demonstrates that \deal\ achieves longer-range energy exploration in KP and MaxCut compared to TSP, as shown in \cref{tab:qnre} (QNRE definition is provided in \cref{med: metrics}).

\section{Discussion}
\label{sec: disc}
% conclusion
% future work

This study presents three contributions to quantum approximation optimization algorithms regarding ansatz encoding and Hamiltonian learning. First, direct parameter passing enhances stability by utilizing each qubit to represent QUBO objective functions by leveraging qubit connectivity from problem definitions. This enables cost Hamiltonian entanglement gates to precisely capture the information from physical qubit layouts. Second, we present experiments across different QPUs demonstrating the \deal\ framework with better noise robustness and higher success rates in solving NP-hard problems compared to previous studies. Third, we develop ZNE-integrated dynamical Hamiltonian learning for \deal\ to mitigate crosstalk errors in quantum hardware using an end-to-end workflow.
% remove related-work and provide a small related topics in intro
% \input{sections/04_related_work}

\section*{Acknowledgements}
% HPCC
% discussion
The authors acknowledge the High Performance Computing Center (HPCC) at Texas Tech University for providing the computational resources that contributed to the research results reported in this paper. URL: http://www.hpcc.ttu.edu. This research used resources of the National Energy Research
Scientific Computing Center, a DOE Office of Science User Facility
This study was supported by the Office of Science of the U.S. Department of Energy
under Contract No. DE-AC02-05CH11231 using NERSC award
NERSC DDR-ERCAP0034486. We thank Jamal Mohammad Khani and Yann Beaujeault-Taudière for their advice on polishing the ideas.

\bibliographystyle{IEEEtran}
\bibliography{bibs/main, bibs/qaoa}

% \newpage
\appendix
\subsection{\deal\ Proof Analysis}
\label{app: qPN}
\begin{figure}[th]
    \centering
    \includegraphics[width=0.75\linewidth]{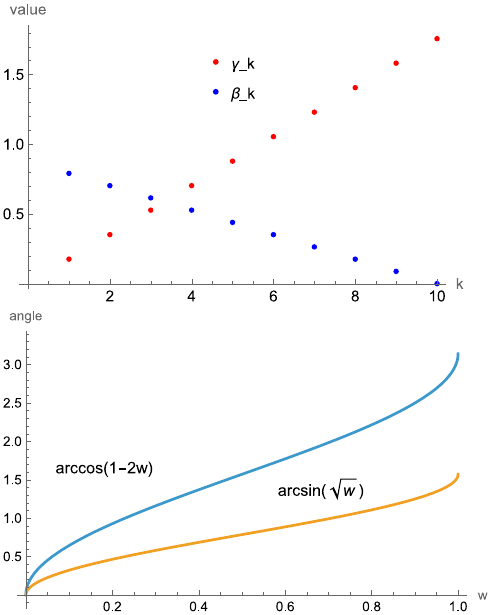}
    \caption{The top panel shows the encoded cost and mixer Hamiltonian values. The bottom panel demonstrates the nonlinear PQN scaling plot corresponding to \cref{eq:cost} and \cref{eq:mixer}.}
    \label{fig:angles2values}
\end{figure}

% example of normalized QUBO matrix
In this section, we analyze the normalization process in the qubit representation, as shown in \cref{fig:med}. We scale the objective function coefficients to the range \([-1,1]\) according to the trigonometric function. While the sin and cos functions for the XY mixer Hamiltonian provide similar observables with differences only in layer initialization, the sin function offers larger differentiation with the cos encoding method in the cost Hamiltonian, as demonstrated in the angle plot in \cref{fig:angles2values}.
We detail the workflow for reproduction as follows:
\begin{enumerate}
    \item 
From the problem graph with nodes $\{0, 1, 2, 3\}$, edge weights $\alpha, \beta, \gamma, \delta$ populate the QUBO matrix elements: $Q_{01} = \alpha$, $Q_{02} = \beta$, $Q_{12} = \gamma$, $Q_{23} = \delta$.

    \item QPN transforms these weights into circuit parameters through cost parameters, where $R_z(\gamma h_i)$ gates receive diagonal parameters $h_i = Q_{ii}$ and $R_z(\gamma J_{ij})$ gates receive off-diagonal elements $J_{ij} = Q_{ij}$. Thus, $J_{01} = \alpha$, $J_{02} = \beta$, $J_{12} = \gamma$, $J_{23} = \delta$.

    \item XY mixer Hamiltonian parameters for $R_{XX+YY}(\beta K_{ij})$ gates receive normalized coupling strengths:
\begin{align}
K_{01} &= \frac{|\alpha|}{|\alpha| + |\beta| + |\gamma| + |\delta|}, \\ K_{02} &= \frac{|\beta|}{|\alpha| + |\beta| + |\gamma| + |\delta|},\\
K_{12} &= \frac{|\gamma|}{|\alpha| + |\beta| + |\gamma| + |\delta|}, \\
K_{23} &= \frac{|\delta|}{|\alpha| + |\beta| + |\gamma| + |\delta|}.
\end{align}

\item Quantum state amplitudes reference the $(p \times n)$ tensors $\Phi^{\gamma}_{k,i}$ and $\Phi^{\beta}_{k,i}$ encoding layer-dependent ($k$) and qubit-dependent ($i$) information from graph weights $\{\alpha, \beta, \gamma, \delta\}$. The initialization follows $\{J_{ij}, K_{ij}\} \rightarrow \{\Phi^{\gamma}_{k,i}, \Phi^{\beta}_{k,i}\} \rightarrow \{\gamma_k^0, \beta_k^0\}$, creating direct mapping from problem structure to quantum circuit parameters.
\end{enumerate}
We conclude that the computational complexity of the \deal\ overhead is $O(n^2)$ with a quantum-sampling cost $O(pS)$, where $S$ denotes the sampling shots. Classical memory requirements are $O(n)$ for storing $w$ and $O(pn)$ for storing $\Phi$.

\subsection{Mapping cases}
\label{app:cases}
In ZNE implementation \(G \mapsto G G^{\dagger} G\), inserted unitary operations increase parameter requirements from QUBO coefficients when hardware \cite{huang2020superconducting, kjaergaard2020superconducting, gambetta2017building, song2019quantum} lacks efficient controlled-operation support (non-local unitary gates). We address three coefficient-compensation cases.

\paragraph*{Case 1: Gates equal QUBO coefficients}
The parameters and qubit terms (unitary operations) are equalized. Single and non-local qubit terms are derived directly from the problem coefficients:
\begin{itemize}
    \item Single-qubit term \( h_i Z_i \) maps to \( RZ(2\gamma h_i) \) gate.
    \item Non-local qubit term \( J_{ij} Z_i Z_j \) maps to \( RZZ(2\gamma J_{ij}) \) gate.
\end{itemize}
where \(\gamma\) controls the cost Hamiltonian evolution and \(J\) represents the QUBO binary variables.

\paragraph*{Case 2: More gates than QUBO coefficients}
In the multi-layer QAOA (\( p > 1 \)), identical QUBO coefficients apply across layers with independent parameters \( \gamma_q \). Hardware constraints and compiler optimizations introduce additional gates owing to connectivity limitations. We employ coefficient duplication for consistent weight application and proportional distribution of repeated terms (e.g., \( J_{ij} \to J_{ij}/2 \) for dual occurrences). Independent parameters \( J_{ij}^{(q)} \) per layer instance enhance flexibility through layer-specific adjustments while preserving the energy landscape.

\paragraph*{Case 3: More QUBO coefficients than gates}
We employ structured approximations and iterative strategies for computational accuracy with efficient resource utilization when redundant coefficients cannot be mapped to variational parameters. Structured approximation enables the truncation of negligible magnitude terms, preventing resource expenditure on minimal contributions. For QUBO matrices with interaction strengths ranging from \(10^{-1}\) to \(10^{-6}\), terms below predefined thresholds (e.g., \(10^{-4}\)) are discarded because of the negligible influence of the optimization landscape. Multi-round encoding distributes coefficients across circuit executions, leveraging temporal redundancy \cite{sundaresan2023demonstrating} as errors are averaged over multiple runs.

\subsection{Bayesian optimization for qubit connectivity}
\label{app: baye}
Given noisy objective function evaluations with residuals $\boldsymbol{r} = \mathbf{f}_{\text{obs}} - \mathbf{f}(\boldsymbol{\lambda})$, we formulate the optimization as minimizing $\mathcal{L}(\boldsymbol{\lambda}) = \frac{1}{2}\|\boldsymbol{r}\|_{\boldsymbol{\Sigma}_{\text{noise}}^{-1}}^2 + \frac{1}{2}\|\boldsymbol{\lambda} - \boldsymbol{\lambda}^{(t)}\|_{\boldsymbol{\Sigma}_{\text{prior}}^{-1}}^2$, where $\boldsymbol{\Sigma}_{\text{noise}} = \sigma_{\text{shot}}^2 \mathbf{I} + \sigma_{\text{gate}}^2 \mathbf{C}_{\text{conn}}$ captures both finite sampling noise and hardware-dependent gate errors scaling with qubit connectivity. Setting $\nabla_{\boldsymbol{\lambda}} \mathcal{L} = -\mathbf{J}^T \boldsymbol{\Sigma}_{\text{noise}}^{-1} \boldsymbol{r} + \boldsymbol{\Sigma}_{\text{prior}}^{-1}(\boldsymbol{\lambda} - \boldsymbol{\lambda}^{(t)}) = 0$ and solving for the parameter update yields the claimed expression, where $\mathbf{J} = \frac{\partial \mathbf{f}}{\partial \boldsymbol{\lambda}}$ is the Jacobian matrix. For implementation, high-dimensional problems (>10 qubits) utilize Gaussian process regression to model $\mathbf{J}$ and handle complex noise correlations, whereas low-dimensional cases employ polynomial approximations enabling closed-form Jacobian computation, ensuring computational efficiency across different problem scales.

\subsection{Benchmark details}
\label{app:benchmark}
\begin{figure*}[tb]
    \centering
    \includegraphics[width=1.0\textwidth]{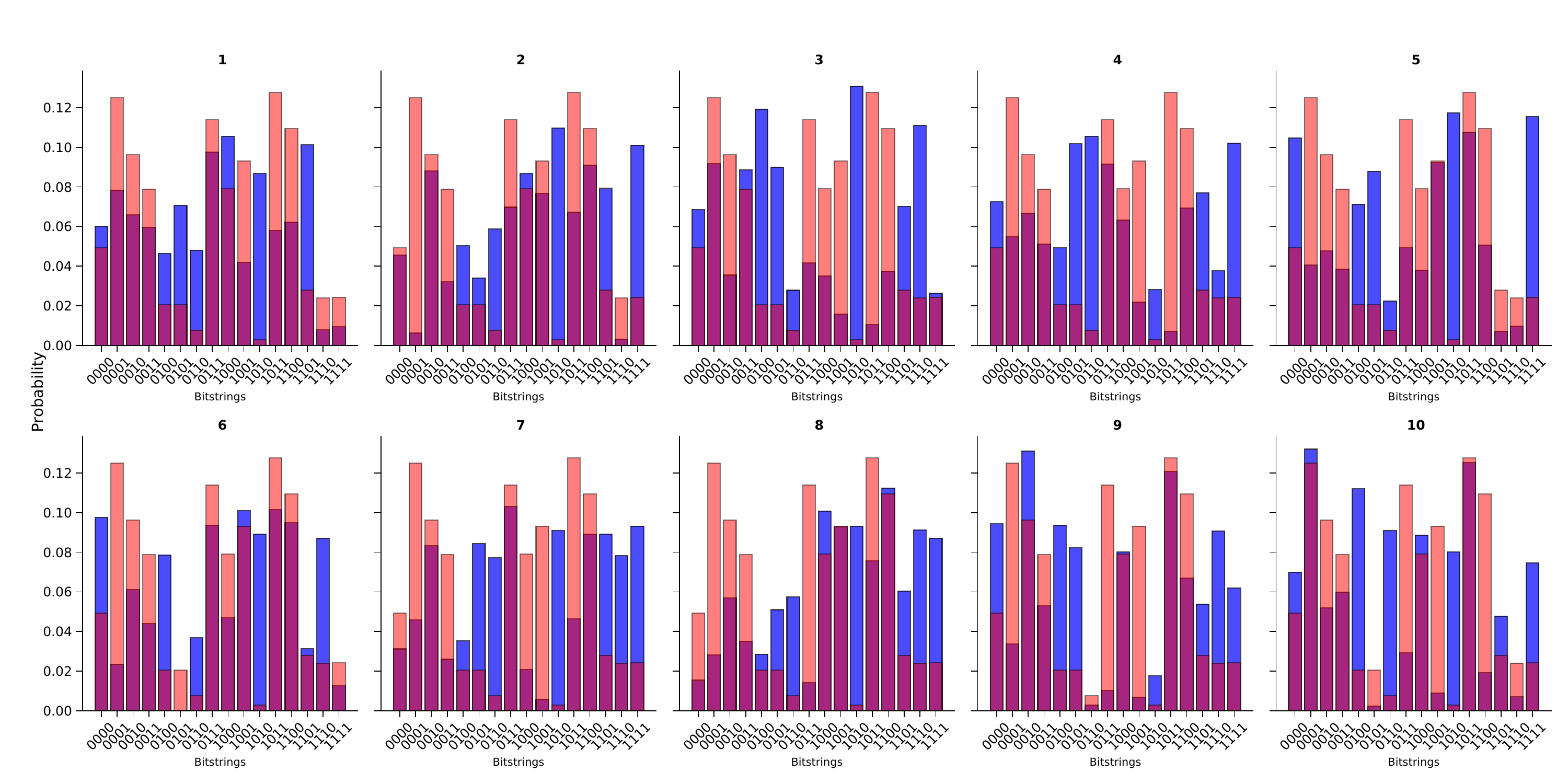}
    
    \vspace{0.5cm}
    
    \includegraphics[width=1.0\textwidth]{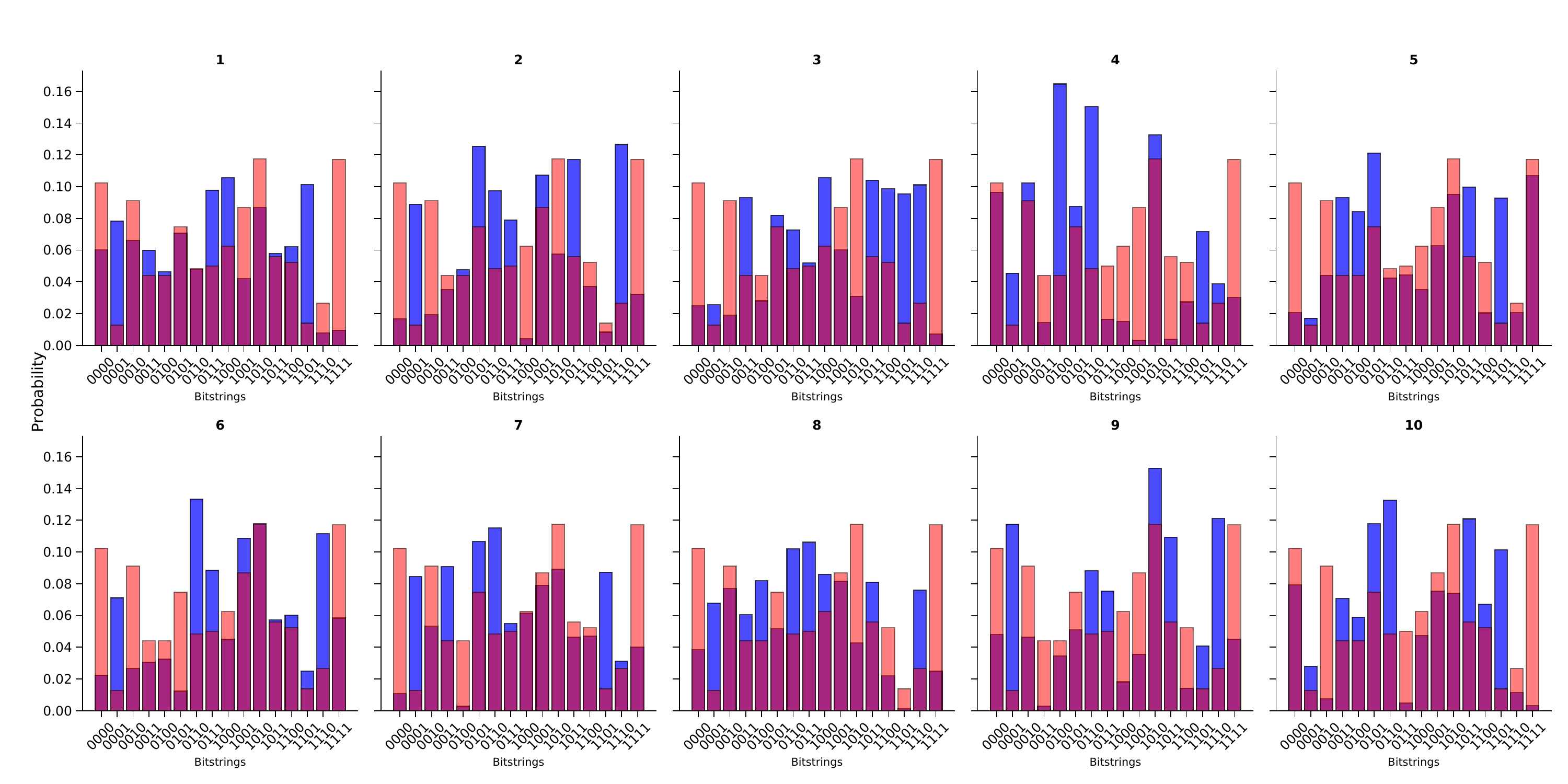}
    
    \caption{Quantum state probability distribution comparison between DEAL (top) and QAOA (bottom) algorithms across 10 circuit layers. The orange and blue histograms represent the Haar-random uniform probability distributions and algorithm-generated probability distributions after optimization, respectively. DEAL achieves maximum single-state probability of 13\% compared to QAOA's 17\%, demonstrating different optimization characteristics.}
    \label{fig:algorithm_comparison_details}
\end{figure*}

\end{document}